\def\Statusstring{}
\documentclass[12pt,letter]{article}

\usepackage{amsmath, amsthm, amssymb}
\usepackage{amssymb,epsfig}
\usepackage{hyperref}
\hypersetup{bookmarks=true}

\def\qed{\hskip 3pt \hbox{\vrule width4pt depth2pt height6pt}}

\newtheorem{Lemma}{Lemma}
\newtheorem{Example}[Lemma]{Example}
\newtheorem{Theorem}[Lemma]{Theorem}

\newtheorem{Definition}[Lemma]{Definition}

\newcommand{\Aut}{\mathop{\mathrm{Aut}}\nolimits}
\newcommand{\Sym}{\mathop{\mathrm{Sym}}\nolimits}

\newcommand{\Cay}{\mathop{\mathrm{Cay}}\nolimits}

\begin{document}

\title{Automorphism groups of graphs \thanks{Based on an Invited Lecture delivered by the author at the {\em Pre-Conference Workshop on Algebraic Graph theory} under the auspices of the {\em 8th Annual Conference of the Academy of Discrete Mathematics and Applications,} Virudhunagar, India, June 2012.}}
\author{Ashwin Ganesan%
  \thanks{Department of Mathematics, Amrita School of Engineering, Amrita University, Amritanagar, Coimbatore - 641~112, India.
  Email: \texttt{ashwin.ganesan@gmail.com}. }}
\date{}

\maketitle

\vspace{-6.5cm}
\begin{flushright}
  \texttt{\Statusstring}\\[1cm]
\end{flushright}
\vspace{+4.3cm}

\begin{abstract}
These lecture notes provide an introduction to automorphism groups of graphs.  Some special families of graphs are then discussed, especially the families of Cayley graphs generated by transposition sets.
\end{abstract}

\bigskip

\section{Introduction}

A {\em permutation} of a set $\Omega$ is a bijection from $\Omega$ to itself.  For example, $\pi = (1,5,4)(3,6)(2) = (1,5,4)(3,6) = [5, 2, 6, 1, 4, 3]$ is a permutation of $\{1,2,3,4,5,6\}$.
We can also write $\pi: 1 \mapsto 5, 2 \mapsto 2, 3 \mapsto 6, 4 \mapsto 1, 5 \mapsto 4, 6 \mapsto 3$.
The inverse of $\pi$ is $\pi^{-1} = (1,4,5)(3,6)(2)$.  $\Sym(\Omega)$ denotes the set of all permutations of $\Omega$.  $S_n$ denotes the symmetric group $\Sym(\{1, \ldots, n\}$.  The composition of permutations is carried out left to right: $(1,2)(2,3) = (1,3,2)$. A {\em transposition} $\tau \in S_n$ is a permutation that interchanges two elements and fixes the remaining elements: $\tau = (i,j)$.

Let $\Gamma:=(V,E)$ be a simple, undirected graph.  An {\em automorphism} of a graph is a permutation of the vertex set that preserves adjacency.  The {\em automorphism group} of $\Gamma$ is the set of permutations of the vertex set that preserve adjacency, i.e., $\Aut(\Gamma):=\{ \pi \in \Sym(V): \pi(E) = E\}$. (Some of the literature uses the notation $E^{\pi}=E$ instead of $\pi(E)=E$.)   Note that if $\pi$ is a bijection from $V$ to iself, then $\pi$ induces a bijection on the 2-subsets of $V$, and so $\pi(E)$ and $E$ have the same cardinality.  So, if $\pi$ is an automorphism, then it necessarily must preserve non-adjacency as well.  So, an equivalent definition is that: $\pi$ is an automorphism of $\Gamma=(V,E)$ if for all $u,v \in V$, $u \sim v$ iff $\pi(u) \sim \pi(v)$. Since the composition of two automorphisms is another automorphism, the set of automorphisms of a graph is a permutation group.

The notation to use $\Gamma$ rather than $G$ for a graph is standard in algebraic graph theory. We shall use $A$ to denote $\Aut(\Gamma)$. Later, we shall use $G$ (and a subset $S \subseteq G$) to denote the group that generates a Cayley graph, i.e $\Gamma=(V,E) = \Cay(G,S)$, and we shall determine its automorphism group $A:=\Aut(\Cay(G,S))$ for some families of $(G,S)$.

\begin{Example}
\upshape \textbf{The n-cycle.}  Let $\Gamma$ be the graph on vertex set $\{1,\ldots,n\}$ having edge set $E := \{ \{1,2\},\{2,3\}, \ldots, \{n,1\}\}$.  The rotation $r = (1,2,\ldots,n)$ is an automorphism of the graph, because $r$ sends $1$ to $2$ and $2$ to $3$, hence $\{1,2\}$ to $\{2,3\}$, and it is clear that $E^r = E$.  Let $s$ be the permutation of $V$ induced by the reflection of the $n$-cycle graph about the axis that passes through the vertex 1 and the center of the $n$-cycle.  Thus, $s=(1)(2,n)(3,n-1) \cdots$.  Note that $\langle r,s \rangle = D_{2n}$, the dihedral group of order $2n$.  So $D_{2n} \le A:=\Aut(\Gamma)$.  We now show that there are no other automorphisms. Let $\pi \in A$.  Suppose $\pi: 1 \mapsto i$.   Then it must send 2 to a neighbor of $i$, i.e. to $i-1$ or $i+1$.  In the first case $\pi$ sends 3 to $i-2$, 4 to $i-3$, and so on.  In the second case, $\pi$ sends 2 to $i+1$ and hence 3 (a neighbor of 2) to a neighbor of $i+1$, i.e. to $i$ or $i+2$, but $\pi:1 \mapsto i$, so $3 \mapsto i+2$.  In either case, $\pi \in D_{2n}$.  Thus $A = D_{2n}$.\qed
\end{Example}

If $\pi \in \Aut(\Gamma)$, then $\pi$ preserves adjacency and hence sends the neighbors of $u$ to the neighbors of $\pi(u)$.  So, if $\pi$ sends $u$ to $v$, then $u$ and $v$ have the same degree (the same number of neighbors).

One can similarly define the automorphism group of $(X,S)$ for any structure $S$ on $X$.  In the case of directed graphs, each element of $S$ would be an ordered pair, and in case of edge-colored digraphs each element of $S$ is an ordered triple.  The automorphism group of $(X,S)$ is the set of permutations $\pi$ in $\Sym(X)$ such that $S^{\pi}=S$.

We now introduce a definition \cite{Harary:1969}:

\begin{Definition}
\upshape Let $A \le \Sym(X), B \le \Sym(Y)$, with $X$ and $Y$ being disjoint.  Their {\em sum} (also known as the product or direct product) $A+B$ is the permutation group $$A+B:= \{(\alpha,\beta): \alpha \in A,\beta \in B\}$$
acting on the disjoint union $X \cup Y$ by the rule:
$$ (\alpha,\beta)(z) =
  \begin{cases}
   \alpha(z) & \text{if } z \in X \\
   \beta(z)     & \text{if } z \in Y.
  \end{cases}
$$
\end{Definition}
So $A+B \le \Sym(X \cup Y)$ and has $|A|~|B|$ elements.

\bigskip It is clear that $\Aut(K_n) = S_n$.

Also, $\Aut(K_{1,n}) = E_1 + S_n \cong S_n$, where $E_1$ is the trivial group.

Consider the graph $K_4$ minus an edge, where $V=\{1,2,3,4\}$ and the edge removed from $K_4$ is $\{2,4\}$ say.  Then $A = \{(1,3),(2,4),(1,3)(2,4),1\}$, and so $A = S_2 + S_2$.

\begin{Example}
\upshape \textbf{The octahedron graph.}  Let $\Gamma$ be the octahedron graph obtained from the solid which is the convex hull of the six points (in 3-dimensional space) $V = \{(\pm1,0,0),(0,\pm1,0),(0,0,\pm1)\}$.
Then, the complement graph of the octahedron graph is $3K_2$.  Also, a graph and its complement have the same automorphism group (this is because every automorphism preserves adjacency as well as non-adjacency).  It is easy to see that the number of automorphisms of $3K_2$ is $2^3 3! = 48$ since every automorphism can move vertices from one edge to another (and there are 3! ways to permute the edges) and on each edge, one can either switch the vertices on that edge or leave them fixed (2 ways for each edge, yielding another $2^3$ automorphisms).  Thus, $|\Aut(\Gamma)|=48$.\qed
\end{Example}

Note that the number of rigid motions of the octahedron is exactly 24.  The transposition that just swaps two nonadjacent vertices of the octahedron graph is an automorphism of the octahedron graph but not a rigid motion of the solid. This is unlike in the case of the $n$-cycle, for which the group of rigid motions of the regular $n$-sided polygon and the group of automorphisms of the $n$-cycle graph are the same.

\begin{Definition} \upshape
The {\em wreath product} (or composition) $A[B]$ of ``A around B'' acts on $X \times Y$.
Suppose $|X|=d$.  Then, every $\pi \in A[B]$ can be written as $\pi = (\alpha; \beta_1,\ldots,\beta_d)$, and $(\alpha; \beta_1, \ldots, \beta_d)(x_i,y_j) = (\alpha x_i, \beta_i y_j)$.
\end{Definition}

The following result is due to Frucht:

\begin{Lemma} \cite{Frucht:1949} The automorphism group of $n$ disjoint copies of a graph $\Gamma$ is $\Aut(n \Gamma) = S_n [\Aut(\Gamma)]$.
\end{Lemma}

It can be seen why Frucht's theorem is true, as we did above for the special case $3K_2$.  In this special case, label the two vertices on the $i$th edge as $(x_i,y_1)$ and $(x_i,y_2)$, for $i=1,2,3$, and we can see that $\alpha \in S_3$ permutes the three edges (i.e. the $x_i$'s), and $\beta_i \in S_2$ (for $i=1,2,3$) acts on the $i$-th edge (i.e. it permutes the $y_i$'s).  So $\Aut(3K_2) = S_3[S_2]$.

\begin{Theorem}
The full automorphism group of the Petersen graph is isomorphic to $S_5$.
\end{Theorem}

\noindent \emph{Proof:}  Let $\Gamma$ be the Petersen graph.  Thus, its vertices are the 2-element subsets of $\{1,2,3,4,5\}$, with two vertices $A$ and $B$ adjacent iff they are disjoint.  We write $ij$ for the vertex $\{i,j\}$.

Every element $\pi \in S_5$ induces a permutation of the 2-element subsets of $\{1,\ldots,5\}$, i.e. induces a permutation $\hat{\pi}$ of $\Gamma$. For example, if $\pi = (1,3,4)(2,5) \in S_5$, then $\hat{\pi}$ sends $\{1,2\}$ to $\{3,5\}$,  and $\hat{\pi} = (12, 35, 42, 15, 32, 45)$ $(13, 34, 41)(25) \in \Sym(V)$.  Furthermore, it is easy to see that different elements of $S_5$ induce distinct permutations of $V$.  And each of these induced permutations is an automorphism of $\Gamma$ because for all $\pi \in S_5$, $A, B \in V$ are disjoint if and only if $\pi(A)$ and $\pi(B)$ are disjoint.  Thus, the map $\phi: S_5 \rightarrow V, \pi \mapsto \hat{\pi}$ is an injective group homomorphism into $\Aut(\Gamma)$, and so $S_5 \cong \phi(S_5) \le \Aut(\Gamma)$.

We now show that $\phi(S_5)$ is the {\em full} automorphism group of $\Gamma$.  It suffices to show that $A \le \phi(S_5)$.  To prove this, let $\pi \in A$.  We show that there exists $\hat{g} \in \phi(S_5)$ such that $\pi \hat{g}=1$.  This would imply that $\pi = \hat{g}^{-1} \in \phi(S_5)$.

Suppose $\pi:\{1,2\} \rightarrow \{a,b\}$.  Let $g_1 \in S_5$ map $ a \mapsto 1, b \mapsto 2$.  Then $\pi \hat{g}_1$ fixes the vertex 12 and hence permutes its neighbors 34, 45 and 35. We consider a few cases:

Case (1) Suppose $\pi \hat{g}_1 $ fixes all three neighbors 34, 45, 35.  So $\pi \hat{g}_1$ permutes the neighbors of 35, and hence fixes 14 and 24, or swaps 14 and 24.

Case (1.1): If $\pi \hat{g}_1: 14 \mapsto 14$, then 15 is sent to a vertex adjacent to 34 and not adjacent to 14, hence 15 is fixed.  Similarly, all the other vertices are seen to be fixed, and so $\pi \hat{g}_1 = 1$, as was to be shown.

Case (1.2):  If $\pi \hat{g}_1: 14 \mapsto 24$.  Then, it swaps 14 and 24.  15 is sent to a neighbor of $24^{\pi \hat{g}_1} = 14$, and hence 15 is sent to 25.  Thus, we see that $\pi \hat{g}_1$ swaps 14 and 24, 15 and 25, and also 13 and 23, and fixes the remaining vertices.  But then $\pi \hat{g}_1 = \hat{g}_2$, where $g_2=(1,2)$.   So $\pi \in \phi(S_5)$.

Case (2): Suppose $\pi \hat{g}_1$ fixes exactly 2 of the 3 vertices 34, 45 and 35.  But this case is not possible, because if it fixes 2 of the 3 vertices, it must also fix the third vertex.

Case (3): Suppose $\pi \hat{g}_1$ fixes exactly 1 of 34, 45 and 35, say 34.  So $\pi \hat{g}_1$ swaps 45 and 35. Let $g_2 = (3,4)$.  Then $\pi \hat{g}_1 \hat{g}_2$ satisfies the conditions of Case (1).

Case (4):  Suppose $\pi \hat{g}_1$ fixes none of 34, 45 and 35.  Say it has $(34,45,35)$ as a 3-cycle.  Let $g_2=(3,4)$.  Then $\pi \hat{g}_1 \hat{g}_2 = (34,35)(45)$, and we are back to Case (3).

Thus, in all cases if $\pi \in A$, then there exists $g_1,\ldots,g_r$ (for some nonnegative integer $r$) such that $\pi \hat{g}_1 \ldots \hat{g}_r = 1$, implying that $\pi \in \phi(S_5)$.
\qed

\bigskip  The family of the generalized Kneser graphs $J(n,k,i)$ is defined as follows.  The vertex set of $J(n,k,i)$ is the set of $k$-element subsets of $\{1,\ldots,n\}$, and two subsets are adjacent iff their intersection has size $i$. Biggs \cite{Biggs:1993} defines the family of Odd graphs $O_k$ to be $J(2k-1,k-1,0)$, a special case of which is the Petersen graph $O_3$.  Thus, the Petersen graph is $J(5,2,0)$ and hence is isomorphic to the complement of the line graph of $K_5$.  It is an open problem to determine the automorphism group of $J(n,k,i)$ for many special cases of these three parameters.  We now give another proof (based on \cite{Lovasz:2007}) for the automorphism group of the Petersen graph.

\bigskip \noindent {\em Second proof:} The automorphism group of the Petersen graph is isomorphic to the automorphism group of the complement of the Petersen graph, which is $L(K_5)$.  Thus, it suffices to find $\Aut(L(K_5))$.  Every automorphism of $K_5$ induces a unique automorphism of $L(K_5)$, as we saw above.  We now show that each automorphism of $L(K_5)$ induce a unique permutation in $S_5$, implying that $\Aut(L(K_5))$ has at most 120 elements.

Let $\pi \in \Aut(L(K_5))$.  Since $L(K_5)$ is a line graph, its 4-cliques correspond to stars $K_{1,4}$ in $K_5$.  Thus, $L(K_5)$ has exactly 5 4-cliques, say $C_1, \ldots, C_5$, where $C_i$ contains the 4 vertices in $L(K_5)$ corresponding to the 4 edges in $K_5$ that are incident to vertex $i$ in $K_5$.  Since $\pi$ is an automorphism, it sends 4-cliques to 4-cliques.  Also, $\pi$ induces a permutation of the $C_i$'s, for if $\pi$ sends two different cliques to the same clique, then $\pi$ is not a bijection on the edge set of $L(K_5)$, which is a contradiction.  Also, distinct automorphisms of $L(K_5)$ induce distinct permutations of the $C_i$'s. (For if we have two distinct automorphisms, say $\pi: 12 \mapsto 34, \pi': 12 \mapsto 35$, then $\{C_1,C_2\}$ is mapped to $\{C_3, C_4\}$ by $\pi$ and to $\{C_3, C_5\}$ by $\pi'$.)
Equivalently, every vertex $ij$ in $L(K_5)$ is uniquely determined by the intersection of two 4-cliques (namely the cliques $C_i$ and $C_j$).  Thus, every automorphism of $L(K_5)$ induces a unique permutation of $\{C_1,\ldots,C_5\}$.
\qed

\begin{Example}
\upshape \textbf{Hypercubes.}   The hypercube graph $Q_n$ has vertex set $V=\{0,1\}^n$, and two vertices are adjacent in this graph iff the two $n$-bit strings differ in exactly one coordinate.  Let $r_z: V \rightarrow V, x \mapsto x+z$ be the translation by $z$.  Here, $x+z$ is the vertex obtained by adding the two bit strings $x$ and $z$ componentwise in the binary field.  Then, $H := \{r_z: z \in V\}$ can be seen to be a group of automorphisms of $Q_n$, and $|H|=2^n$.  Also, let $K$ be the group of permutations of the vertex set induced by the permutation of the $n$ coordinates of the bit strings.  Then $K$ fixes $00 \ldots 0$, and $|K|=n!$. Also, $H \cap K$ is trivial, and so $|HK| = |H|~|K|/~|H \cap K| = 2^n n!$.  It can be shown that $|A| \le 2^n n!$, so that $HK$ is the full automorphism group.  To show $|A| \le 2^n n!$, note that any automorphism that fixes the vertex $00 \ldots 0$ and each of its neighbors in $Q_n$ also fixes all the remaining vertices. Thus, $|A_e| \le n!$, where $A_e$ is set of the automorphisms that fix the vertex $e=00 \ldots 0$. By the orbit-stabilizer lemma, $|A| = |A_e| |V|$,  and we get the desired bound. It is shown in \cite{Harary:2000} that $\Aut(Q_n)$ is isomorphic to the wreath product $S_n[S_2]$.\qed
\end{Example}

\bigskip In algebraic graph theory, open-source software packages such as GAP and SAGE can be helpful with suggesting or testing conjectures.  For example, an open problem is to determine the diameter of the modified-bubble sort graph $MBS(n)$ (cf. \cite{Jerrum:1985}), and it can be confirmed with the help of a computer that for $n=3,\ldots,10$, the diameter of $MBS(n)$ is equal to $\lfloor n^2 /4 \rfloor$. (When the sequence of values obtained using the computer is entered into the Online Encyclopedia of Integer Sequences, this formula is displayed; this formula is also mentioned in \cite{Heydemann:1997}).  No proof of this formula for all $n$ is available in the literature.

\section{Cayley graphs generated by transposition sets}

A very promising area of research is Cayley graphs generated by transposition sets.  I shall now mention some results related to the automorphism groups of these families of graphs.

Let $G$ be a group and $S$ a subset of $G$. The Cayley digraph (also known as the Cayley diagram \cite{Bollobas:1998}) of $G$ with respect to $S$, denoted by $\Cay(G,S)$, is the digraph with vertex set $G$ and with an arc from $g$ to $sg$ whenever $g \in G$ and $s \in S$.  When $S$ is closed under inverses (i.e. when $S^{-1} := \{s^{-1}: s \in S\} = S$), $(g,h)$ is an arc of the Cayley digraph if and only if $(h,g)$ is an arc, and so we can identify the two arcs $(g,h)$ and $(h,g)$ with the undirected edge $\{g,h\}$.  When $1 \notin S$, the digraph contains no self-loops.  Thus, when $1 \notin S=S^{-1}$, we view the Cayley graph $\Cay(G,S)$ as a simple, undirected graph. $\Cay(G,S)$ is connected if and only if $S$ generates $G$.

For a group $G$, define the map of right translation by $z$, $r_z: G \rightarrow G, g \mapsto gz$.  The right regular representation of $G$ is the permutation group $R(G):=\{r_z: z \in G\}$, and $R(G)$ is isomorphic to $G$.  The automorphism group of a Cayley graph $\Gamma := \Cay(G,S)$ contains the right regular representation $R(G)$, and hence all Cayley graphs are vertex-transitive (cf. \cite{Biggs:1993}). Since $R(G)$ is regular, $\Aut(\Gamma) = \Aut(\Gamma)_e R(G)$, where $\Aut(\Gamma)_e$ is the stabilizer of $e$ in $\Aut(\Gamma)$.  The set of automorphisms of the group $G$ that fix $S$ setwise, denoted by $\Aut(G,S):= \{ \pi \in \Aut(G): S^{\pi}=S\}$, is a subgroup of $\Aut(\Gamma)_e$ (cf. \cite{Biggs:1993}).  For any Cayley graph $\Gamma:=\Cay(G,S)$, the normalizer $N_{\Aut(\Gamma)} R(G)$ is equal to $R(G) \rtimes \Aut(G,S)$ (cf. \cite{Godsil:1981}), where $\rtimes$ denotes the semidirect product (cf. \cite{Dixon:Mortimer:1993}).  A Cayley graph $\Gamma:=\Cay(G,S)$ is said to be {\em normal} if $R(G)$ is a normal subgroup of $\Aut(\Gamma)$, or equivalently, if $\Aut(\Gamma) = R(G) \rtimes \Aut(G,S)$.

Let $S$ be a set of transpositions in $S_n$.  The transposition graph $T(S)$ of $S$ is defined to be the graph with vertex set $\{1,2,\ldots,n\}$ and with $\{i,j\}$ an edge of $T(S)$ whenever $(i,j) \in S$.  A set of transpositions of $\{1,\ldots,n\}$ generates $S_n$ if and only if the transposition graph $T(S)$ is connected, and a set of transpositions is a minimal generating set for $S_n$ if and only if $T(S)$ is a tree (cf. \cite{Godsil:Royle:2001}).  A set of transpositions whose transposition graph is a tree is called a transposition tree.  Cayley graphs of permutation groups generated by transposition sets have been well studied, especially for consideration as the topology of interconnection networks \cite{Lakshmivarahan:etal:1993}, \cite{Heydemann:1997}.

Some particular families of Cayley graphs $\Cay(Gr(S),S)$ are defined as follows:
\begin{itemize}
\item If $T(S) = K_{1,n-1}$, then $\Cay(S_n,S)$ is called the star graph.
\item If $T(S) = P_n$, then $\Cay(S_n,S)$ is called the bubble-sort graph.
\item If $T(S) = C_n$, then $\Cay(S_n,S)$ is called the modified bubble-sort graph.
\item If $T(S) = nK_2$, then $\Cay(Gr(S),S)$ is the hypercube.
\item If $T(S) = K_n$, then $\Cay(S_n,S)$ is the complete transposition graph.
\item If $T(S) = K_{k,n-k}$, then $\Cay(S_n,S)$ is the generalized star graph.
\end{itemize}

We now mention some results, for which the proofs can be found in the given references.

\begin{Lemma} \cite{Godsil:Royle:2001}
Let $S$ be a set of transpositions of $\{1,\ldots,n\}$.  Then, $S$ generates $S_n$ if and only if the transposition graph $T(S)$ is connected.
\end{Lemma}

\bigskip If $\pi \in S_n$ is a permutation and $i$ and $j$ lie in different cycles of $\pi$, then the number of cycles in the product of $\pi$ and $(i,j)$ is exactly one less than the number of cycles in $\pi$ (cf. \cite{Biggs:2003}).  It can be shown that a product of $n-1$ transpositions is an $n$-cycle if and only if the corresponding $n-1$ edges form a tree in the transposition graph.

\begin{Lemma}\cite{Godsil:Royle:2001}
 Let $S$ be a set of transpositions of $\{1,\ldots,n\}$.  Then the following statements are equivalent:

(1) $S$ is a minimal generating set for $S_n$

(2) $T(S)$ is a tree

(3) The product of the elements of $S$ in any order is a cycle of length $n$.
\end{Lemma}

\begin{Lemma} \cite[Lemma 3.10.3]{Godsil:Royle:2001}  \label{GodsilRoyle:4cycles}
Let $S$ be a set of transpositions such that the transposition graph $T(S)$ does not contain a triangle, and let $t,k \in S$. Then, $tk = kt$ if and only if there is a unique 4-cycle in $\Cay(S_n,S)$ containing $e, t$ and $k$.
\end{Lemma}

Among the Cayley graphs of the symmetric group generated by a set of transpositions of $\{1,\ldots,n\}$, the automorphism groups of the following Cayley graphs are known.
When $T(S)$ is a star, the corresponding Cayley graph is called a star graph and has automorphism group $S_{n-1} R(S_n)$ \cite{Huang:Zhang:STn:submitted}.  The Cayley graph generated by a path graph is called a bubble-sort graph and has automorphism group isomorphic to $Z_2 R(S_n)$ \cite{Zhang:Huang:2005}.  The automorphism group of the Cayley graph generated by an asymmetric transposition tree is $R(S_n)$  and hence is isomorphic to $S_n$ \cite{Godsil:Royle:2001}.  The automorphism group of the Cayley graph $\Cay(S_n,S)$ generated by an arbitrary transposition tree $T(S)$ is shown in \cite{Feng:2006} to be isomorphic to $R(S_n) \rtimes \Aut(S_n,S)$.  All these Cayley graphs are also known to be normal.

\begin{Theorem} \cite{Feng:2006}
Let $S$ be a set of transpositions generating $S_n$. Then, $\Aut(S_n,S) \cong \Aut(T(S))$.
\end{Theorem}

\begin{Theorem} \label{lemma:feng:normalTS}  \cite[Corollary 2.5]{Feng:2006}
Let $S$ be a set of transpositions generating $S_n$ satisfying the following condition for any two distinct transpositions $t,k \in S$: $tk = kt$ if and only if $\Cay(S_n,S)$ has a unique 4-cycle containing $e,t$ and $k$, and if $tk \ne kt$ then $\Cay(S_n,S)$ has a unique 6-cycle containing $e,t,k$ and a vertex at distance 3 from $e$.  Then $\Cay(S_n,S)$ is normal and has automorphism group isomorphic to $R(S_n) \rtimes \Aut(S_n,S)$.
\end{Theorem}

The modified bubble-sort graph $MBS(n)$ is the Cayley graph of $S_n$ with respect to the set $S$ of cyclically adjacent transpositions $\{ (1,2),(2,3),\ldots,(n,1)\}$.  Thus, $T(S) = C_n$ in this case.

The results that were mentioned above on the automorphism groups of Cayley graphs generated by transposition graphs that are  particular trees, asymmetric trees, and arbitrary trees have been generalized in \cite{Ganesan:arXiv:autMBSn:May2012} to arbitrary transposition graphs that are connected and do not contain triangles or 4-cycles:

\begin{Theorem} \cite{Ganesan:arXiv:autMBSn:May2012}
Let $MBS(n):=\Cay(S_n,S)$ be the modified bubble-sort graph.  Let $t,k \in S$.  If $tk \ne kt$, then the modified bubble-sort graph $MBS(4)$  does not contain a unique 6-cycle passing through $e, t, k$ and a vertex at distance 3 from $e$, and $MBS(n)$ does contain a unique 6-cycle passing through $e, t, k$ and a vertex at distance 3 from $e$ for all $n \ge 5$.
\end{Theorem}

\begin{Lemma} \cite{Ganesan:arXiv:autMBSn:May2012}
If $s_i,s_{i+1} \in S$ are two adjacent transpositions in $T(S) = C_4$, then there are exactly eight distinct 6-cycles in $MBS(4)$ that contain $e, s_i, s_{i+1}$ and a vertex at distance 3 from $e$.  The total number of vertices in 6-cycles of $MBS(4)$ that contain $e, s_i$ and $s_{i+1}$ and that are at distance 3 from $e$ is exactly 6.
\end{Lemma}

\begin{Theorem} \cite{Ganesan:arXiv:autMBSn:May2012}
The automorphism group of the modified bubble-sort graph $MBS(n)$ is isomorphic to $D_{2n} V_4 R(S_n)$, if $n=4$, and is isomorphic to $D_{2n} R(S_n)$, if $n \ge 5$. Here, $D_{2n}$ is the dihedral group of order $2n$, $V_4$ is the Klein four-group, and $R(S_n)$ is the right regular representation of $S_n$.
\end{Theorem}

\begin{Theorem} \cite{Ganesan:arXiv:autMBSn:May2012}
The modified bubble-sort graph $MBS(n)$ is not normal if $n=4$ and is normal if $n \ge 5$.
\end{Theorem}

\begin{Theorem} \cite{Ganesan:arXiv:autMBSn:May2012}
Let $S$ be a set of transpositions generating $S_n$, let $T(S)$ denote the transposition graph, and let $\Gamma:=\Cay(S_n,S)$.  Let $k, t \in S$ be distinct transpositions.  If $tk \ne kt$ and $T(S)$ does not contain triangles or 4-cycles, then there is a unique 6-cycle in $\Gamma$ containing $e,t,k$ and a vertex at distance 3 from $e$.  Moreover, if $T(S)$ does contain a triangle or a 4-cycle, then there exist $t,k \in S$ such that $tk \ne kt$ and such that there does not exist a unique 6-cycle in $\Gamma$ containing $e, t, k$ and a vertex at distance 3 from $e$.
\end{Theorem}

\begin{Theorem} \cite{Ganesan:arXiv:autMBSn:May2012} Let $S$ be a set of transpositions generating $S_n$ such that the transposition graph $T(S)$ does not contain triangles or 4-cycles.  Then, $\Cay(S_n,S)$ is normal and has automorphism group isomorphic to $R(S_n) \rtimes \Aut(S_n,S)$.
\end{Theorem}

We summarize these results on the automorphism group $A:=\Aut(\Cay(S_n,S))$ as follows:
\begin{itemize}
\item If $T(S)$ is a star, $A \cong S_{n-1} R(S_n)$ \cite{Huang:Zhang:STn:submitted}
\item  If $T(S)$ is a path, $A \cong Z_2 R(S_n)$ \cite{Huang:Zhang:STn:submitted}
\item If $T(S)$ is an asymmetric tree, $A \cong S_n$ \cite{Godsil:Royle:2001}
\item If $T(S)$ is a tree, $A \cong  R(S_n) \rtimes \Aut(S_n,S)$ \cite{Feng:2006}
\item If $T(S)$ is an $n$-cycle, $A \cong D_{2n} V R(S_n)$ if $n=4$ and $A \cong D_{2n}R(S_n)$ if $n \ge 5$ \cite{Ganesan:arXiv:autMBSn:May2012}
\item If $T(S)$ is connected and does not contain triangles or 4-cycles, then $A \cong  R(S_n) \rtimes \Aut(S_n,S)$ \cite{Ganesan:arXiv:autMBSn:May2012}
\end{itemize}

\section{Other graph invariants}

A very promising area of research is Cayley graphs generated by transposition sets, as was mentioned earlier.  Besides the automorphism group, other properties of interest are: diameter, vertex connectivity, fault tolerance, indices of routings, diameter vulnerability, and distributed routing algorithms, among others.  See  \cite{Ganesan:arXiv:effalg:May2012} and the references therein for open problems and questions on the diameter of these families of Cayley graphs; see also the survey articles \cite{Lakshmivarahan:etal:1993}, \cite{Heydemann:1997}.  Many of these problems have applications in interconnection networks.

\bigskip The rich theory of hypercubes can be generalized to or investigated for other Cayley graphs.  For example, when the transposition graph $T(S)$ belongs to a particular family of graphs - say trees, complete bipartite graphs, $n$ disjoint copies of some simple graph, cycles, etc - much remains to be investigated.

\bibliographystyle{plain}
\bibliography{refsaut}

\end{document}